\documentclass[fleqn,10pt]{wlscirep}

\usepackage{amsmath}
\usepackage{amssymb}
\usepackage{graphicx}
\usepackage{caption}
\usepackage{subcaption}

\title{Fluid leakage near the percolation threshold}

\author[1]{Wolf B. Dapp}
\author[1,2,*]{Martin H. M\"user}

\affil[1]{Forschungszentrum J\"ulich,
John von Neumann Institut f\"ur Computing and
J\"ulich Supercomputing Centre, 52425 J\"ulich, Germany}
\affil[2]{Universit\"at des Saarlandes, Lehrstuhl f\"ur Materialsimulation, Saarbr\"ucken, Germany}
\affil[*]{m.mueser@fz-juelich.de}

\begin{abstract}
{
Percolation is a concept widely used in many fields of research and 
refers to the propagation of substances through porous media (e.g., 
coffee filtering), or the behaviour of complex networks (e.g., 
spreading of diseases). 
Percolation theory asserts that most percolative processes are universal, 
that is, the emergent powerlaws only depend on the general, statistical 
features of the macroscopic system, but not on specific details of the 
random realisation. 
In contrast, our computer simulations of the leakage through a seal---applying 
common assumptions of elasticity, contact mechanics, and fluid dynamics---show that 
the critical behaviour (how the flow ceases near the sealing point) 
solely depends on the microscopic details of the last constriction.
It appears fundamentally impossible to accurately predict from 
statistical properties of the surfaces alone how strongly we have 
to tighten a water tap to make it stop dripping and also how it starts 
dripping once we loosen it again.
}
\end{abstract}

\begin{document}

\flushbottom
\maketitle

\thispagestyle{empty}
\section*{Introduction}

Seals and gaskets are crucial components in many hydraulic systems
such as water taps, pipes, pumps, or valves\cite{Flitney07}.
Their main function is to prevent undesired or uncontrolled leakage of
gases or fluids  from one region to another. 
Despite their importance, attempts to estimate leak rates of seal systems 
from first principles succeeded for the first time only less than a decade 
ago\cite{Persson08JPCM,Lorenz09EPL,Lorenz10EPJE}. 
Earlier treatments could not accurately predict the distribution of microscopic 
interfacial separations in a mechanical contact, which is needed for the 
fluid-mechanics aspect of the problem. Persson's contact mechanics 
theory\cite{Persson01,Persson05JPCM} provides this information and computes
the leakage in terms of Bruggeman's effective-medium 
approximation\cite{Bruggeman35} to Reynolds equation\cite{Reynolds86}.

In a previous publication\cite{Dapp12PRL}, we demonstrated that Persson's contact 
mechanics theory combined with a slightly modified version of Bruggeman's 
effective-medium approximation reproduced almost perfectly the results of computer 
simulations, in which an ideally well-defined leakage problem was solved to high 
numerical precision. 
The favourable comparison of theory and simulations benefited to some degree from 
fortuitous error cancellation:  Persson theory slightly underestimates the rate at 
which (mean) gaps diminish with increasing load,
which almost exactly compensates the minor
overestimation of leakage in Bruggeman's approximation.
Persson's treatment is therefore certainly accurate enough to explain why leakage 
through interfaces decreases roughly exponentially with the mechanical 
load\cite{Lorenz09EPL,Lorenz10EPJE,Armand64} pressing two (elastic) bodies 
against each other, where at least one of them has a 
self-affine rough surface (see the method section).

Although Persson theory has proven successful in describing leakage over a broad
parameter range, one cannot expect it to hold near the sealing transition. 
One reason is that  mean-field theories like Bruggeman's, which is part of 
Persson's approach to leakage, are known to fail near critical points, even if 
they perform quite well outside the critical region\cite{Kirkpatrick73}. 
Alternative, percolation-theory-based treatments of 
leakage\cite{Bottiglione09,Bottiglione09TI} or related approaches assuming that 
most of the fluid pressure drops near a single, narrow constriction (or a 
two-dimensional network of constrictions)\cite{Lorenz09EPL} also risk to fail 
in the vicinity of the sealing transition.
This is because length and width of an isolated
constriction show different scaling with the applied load\cite{Dapp15EPL} in 
contrast to assumptions made in the respective theories. 

In this work, we investigate the fluid leakage through a mechano-hydraulic interface
by means of computer simulations.
In contrast to previous studies\cite{Dapp12PRL,Vallet09a,Vallet09b} our focus lies 
on calculating leakage between randomly rough bodies near the percolation threshold.
A particular motivation to revisit the problem stems from our observation that 
local details, such as the presence or absence of adhesion between the surfaces, 
affect the conductance exponent of isolated constrictions\cite{Dapp15EPL}. 
It remains unclear if or to what degree the critical behaviour (evaluated near but 
not too close to the percolation threshold) is determined by the disorder at large 
length scales, which is usually considered central in percolation 
theory\cite{Stauffer91}.

\section*{Results and Discussion} 

\subsection*{Adhesion-free sealing transition in the continuum limit}
We begin the analysis of leakage near the percolation threshold by simulating our
``default model''. 
It is based on approximations that are commonly made to study either the 
contact-mechanics or the fluid-mechanics aspects of our leakage problem:
self-similar surface roughness, linearly elastic bodies,  small surface slopes, 
and absence of adhesion between the surfaces.
Fluid flow through the interface is treated in terms of the Reynolds equation. 
Some of the approximations of our default model are relaxed below. 
More details are given in the method section.

\begin{figure*}[htbp]
\centering
\includegraphics[width=1.0\linewidth]{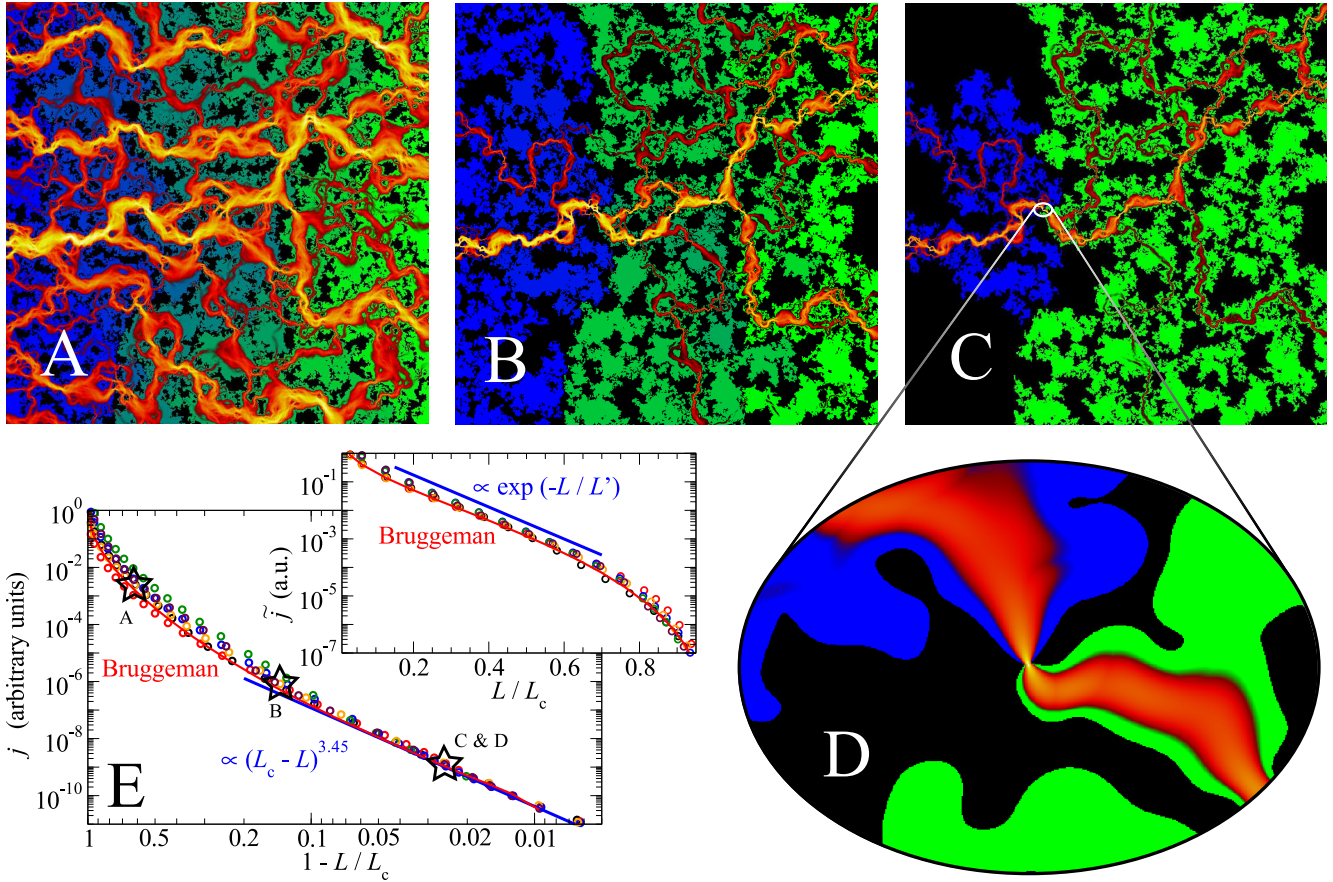}
\caption{\label{fig:defaultSystem}
A--D: 
Visualisation of the fluid flow through a microscopically rough contact at different 
loads, as indicated in panel E. 
Black colour marks regions that do not belong to the percolating fluid channel.
Blue and green colours indicate the fluid pressure, which drops from one (blue)
on the left-hand side of the interface to zero (green) on the right-hand side. 
Red and yellow indicate the absolute value of the fluid current density. 
E main panel: Double-logarithmic representation of the mean leakage current $j$ 
as a function of the reduced load $1-L/L_{\rm c}$. 
Differently coloured symbols represent different random realisations of the surface 
roughness. 
Data is shifted vertically (by as much as a factor of 10) to superimpose in the 
critical region. 
In the inset of panel E, the dimensionless load $L/L_{\rm c}$ is plotted linearly 
and the current is now normalised (shifting factors $\lesssim 2$) such that it 
superimposes in the domain where it decreases exponentially with load.
Red lines show the predictions of effective medium theory, modified such 
that it reproduces the exact critical load for a given random realisation 
(see Ref.\citenum{Dapp12PRL}).
}
\end{figure*}

Leakage flow for our default system is shown in Fig.~\ref{fig:defaultSystem}
for different reduced loads $1-L/L_{\rm c}$, 
where $L$ is the absolute load squeezing the surfaces together and $L_{\rm c}$ 
is the critical load, defined as the largest load at which at least one fluid  
channel still percolates from the right to the left side of the interface. 
Our data is based on different surfaces, which are produced with identical 
stochastic rules but different random seeds. 
To enhance sampling, we also considered inverted and 90$^\circ$ rotated surfaces.
All realisations show similar behaviour: 
For very small loads,  the current decreases very quickly
with increasing load before the dependence becomes roughly exponential. 
At $1-L/L_{\rm c} = O(10\%)$, a crossover to a powerlaw ensues 
\begin{equation}
j \propto (1-L/L_{\rm c})^\beta,
\label{eq:criticalEq}
\end{equation}
where $\beta$ is the conductance exponent. 
The value of $\beta$ deduced from the data is consistent with the one we identified 
for isolated, 
single-wavelength
constrictions\cite{Dapp15EPL}, i.e., $\beta=69/20$. 
This value is much greater than typical conductance exponents for seemingly
related percolation problems such as the two-dimensional
random (on/off) resistance network, for which $\beta = 1$\cite{Webman77}.
Surprisingly, Bruggeman's effective medium approach predicts the current
quite accurately for most random surface realisations investigated in this study,
even close to the percolation point and in all cases does it find cross-over
loads within roughly 10\% percent at which the exponential load-current relation 
ceases to be valid.

To rationalise how the mean fluid flow develops as a function of load, 
it is instructive to visualise the spatially resolved fluid pressure and current 
density for a particular random realisation.
This is done in Fig.~\ref{fig:defaultSystem}~A--D.
In the domain where flow decreases exponentially with load, 
the fluid pressure drops in a quasi-discrete fashion at a number of constrictions.
These constrictions are distributed seemingly randomly throughout the interface
thereby roughly mimicking the conditions assumed in the derivation of 
Bruggeman's effective-medium theory. 
Once the fluid pressure drops predominantly at a single constriction,
see Fig.~~\ref{fig:defaultSystem}~C and D, 
mean-field theory may still be correct, albeit only incidentally. 
In the language of percolation theory, all current now goes through one
hot bond.
In contrast to assumptions commonly made for random disorder, 
the resistivity assigned to individual points 
is not discrete but it changes continuously with the control parameter and 
eventually diverges at the critical point.
In our case, the control parameter is the load, while in most percolation models
it would be the probability with which a bond (or a vertex point or an individual
point in a continuous domain) 
would be assigned a (fixed) finite or infinite resistance. 

Our system can be characterised as having correlated disorder\cite{Schrenk13}
(if the gap is large at a given position, then the gap is also large nearby) 
and at the same time long-range interactions\cite{Grassberger13}
(the elastic Green's functions in real space decay with $1/r$). 
Apparently, the way in which these two ingredients are combined here
turns percolation of seals into a local problem such that it is not possible to assign
a (unique) universality class to the leakage problem, even if the stochastic
properties of the problem are fully defined. 
We note that neither changing the Hurst exponent nor increasing system size alters
the observed behaviour in a qualitative fashion.
While increasing the range over which the surface spectrum is self similar can and 
does affect the low-load flow quite dramatically, the critical region does
not appear to be affected, at least not for practically relevant spectra,
in which self-similarity is rarely observed for more than five or six decades 
in wavelength. 
In all cases we find that critical behaviour, i.e., the range of loads in which 
equation~(\ref{eq:criticalEq}) holds to within a few ten percents, starts to set 
in at roughly 0.8 to 0.9 times the critical load. 
We substantiated these claims by running additional 
simulations for $H = 0.3$,
by extending the ratio of roll-off wavelength $\lambda_{\rm r}$ and short
wavelength cutoff $\lambda_{\rm s}$ from 64 to 256 (and within
Persson theory to $10^7$),
and by extending the ratio of system size ${\cal L}$ and $\lambda_{\rm r}$ 
from 2 to 16. 
\subsubsection*{Size-dependence of the critical regime}

In the critical regime, the pressure drops predominantly at a single constriction.
One might argue that large systems have a smaller critical regime, because significant
pressure drops can then occur at several constrictions.
We therefore analysed how the size of the critical regime depends on the system size.
For this, we changed the ratios $\epsilon_{\rm t} = {\cal L}/\lambda_{\rm r}$ and 
$\epsilon_{\rm f} = \lambda_{\rm r}/\lambda_{\rm s}$.
Here, we may associate $\epsilon_{\rm t} \to 0$ with the thermodynamic limit
and $\epsilon_{\rm f} \to 0$ with the fractal limit.
In real applications, the true (mathematical) limits have no significance, which is why
we content ourselves with projections of our results to more realistic values of
$\epsilon_{\rm t} = \mathcal{O}(10^{-1})$ and $\epsilon_{\rm f} = \mathcal{O}(10^{-5})$. 

\begin{figure}[htbp]
\centering
\includegraphics[width=0.95\linewidth]{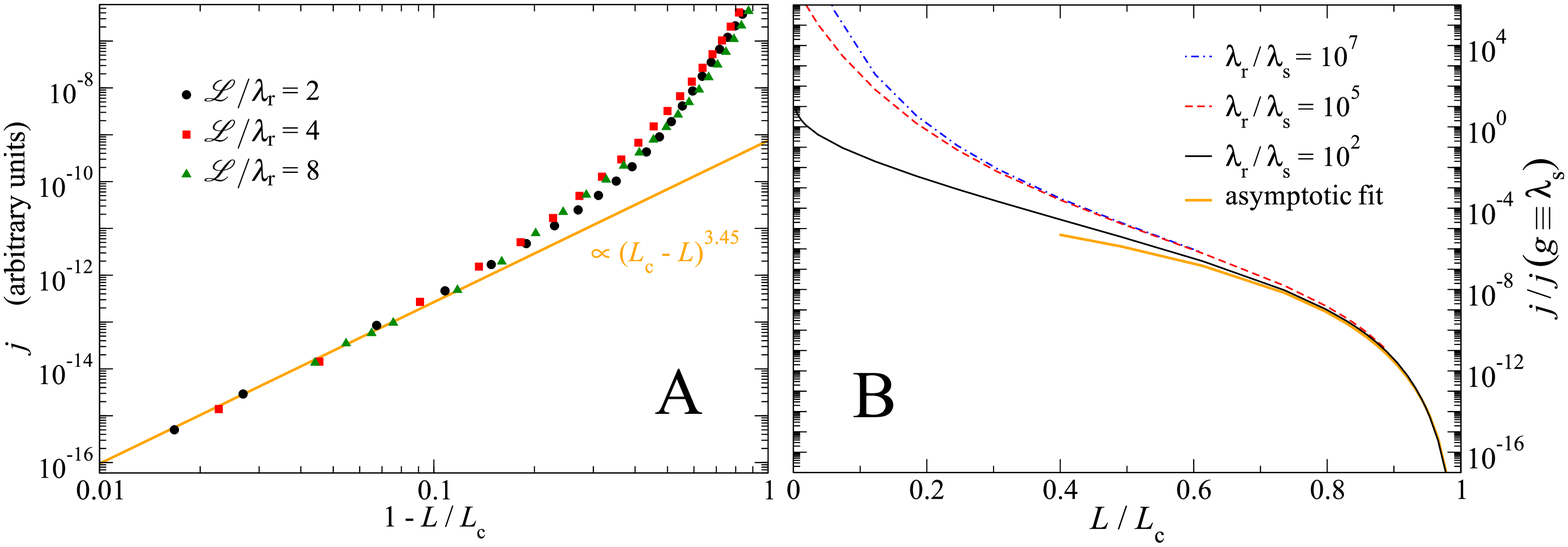}
\caption{\label{fig:sizeDependence}
Size-dependence of the critical regime. {\bf Panel A:} Fluid current $j$ as a 
function of reduced load for different ratios of system size and rolloff wavelength 
${\cal L}/\lambda_{\rm r}$. The current is shifted vertically to superimpose the data 
close to the percolation threshold. Other dimensionless quotients are kept constant, 
e.g., $\lambda_{\rm r}/\lambda_{\rm s} = 64$  and $\lambda_{\rm s}/a = 16$. 
{\bf Panel B:} Fluid current $j$ in Persson theory as a function of the normalised 
load $L/L_{\rm c}$ for different ratios of roll-off wavelength and short-wavelength 
cutoffs, $\lambda_{\rm r}/\lambda_{\rm s}$. As the normalising factor for $j$, we 
chose the current that one would obtain if the gap $g$ were set to $\lambda_{\rm s}$ 
everywhere. 
}
\end{figure}

Figure~\ref{fig:sizeDependence}A reveals that changing $\epsilon_{\rm t}$ does not 
have a sufficiently strong systematic effect to dominate the fluctuations between
different random realisations,  at least not when changing $\epsilon_{\rm t}$ by a
factor of four, i.e., for our three choices of $\epsilon_{\rm t}$ there is not even
a monotonic trend. 
We note that we carried out a disorder average for $\epsilon_{\rm t} = 1/2$ over
16 different realisations but considered the large system $\epsilon_{\rm t} = 1/8$
sufficiently self-averaging. 
Since $\epsilon_{\rm t}$ is never a very small number in practice, we conclude that
the critical leakage regime in real applications should not be much reduced in size
compared to the calculations presented here. 

We also changed $\epsilon_{\rm f}$  and trends on the size of the critical regime were again 
difficult to ascertain, due to large statistical scatter.
We therefore considered realistic values for $\epsilon_{\rm f}$ within Persson 
theory, where we proceed for the calculation of the gap distribution function
as in Ref.~\citenum{Almqvist11}, 
and present our results on the mean flow for $\epsilon_{\rm f} = 10^{-2}$ 
and $10^{-5}$ as well as a rather small value (irrelevant for practical
applications) of $10^{-7}$, 
see Fig.~\ref{fig:sizeDependence}B.
The results for the analysed values of $\epsilon_{\rm t}$ 
are very close at loads approaching the critical load. 
While the critical region is slightly reduced for small values of 
$\epsilon_{\rm f}$, it is clearly revealed that decreasing $\epsilon_{\rm f}$ below
$10^{-5}$ has only marginal effects at loads exceeding $L_{\rm c}/2$. 
This could have been expected from the following argument:
decreasing $\epsilon_{\rm f}$ corresponds to adding roughness at long wavelengths.
Since the effective elastic compliance decreases with the inverse wavelength, 
this extra roughness is immediately accommodated by the elastic seal. 
We conclude that from a mathematical point of view the expected flow or conductivity has a 
fractal limit, which is only (approximately) reached in practice for loads not 
too small compared to the critical load.

\subsection*{Critical leakage for negative slip lengths}

Conductance exponents in percolation theory frequently turn out to be universal,
that is, they remain unaltered when details of a model change.
In view of this finding, we explore whether the conductance exponent
also remains unaltered with small alterations to the default leakage model. 
One simple modification is to assume that the fluid flow velocity does not 
extrapolate to zero precisely at the walls (so-called stick condition) but 
already a distance $d_0$ before the wall.
In fluid dynamics, $d_0$ is called a (negative) slip length.
In a recent work\cite{SchnyderEtAl2015}, softening the fluid-obstacle repulsion, 
in effect using a positive sliplength, suppressed the expected, 
universal critical behaviour for 2D particle transport through porous media, albeit 
for non-correlated obstacles.

In the present context, one could argue that a negative slip length accounts
for the finite size of particles to lowest order:
the fluid particles can only penetrate gaps with a height greater than $2\,d_0$.
The local fluid conductivity now scales with $\{g(x,y)-2d_0\}^3$ rather than
with $g^3(x,y)$, where $g(x,y)$ is the gap as deduced from the contact-mechanics
calculation at an in-plane coordinate $(x,y)$. 
To analyse the effect of finite, negative slip lengths, we solve the 
Reynolds equation for the same gap topography as before, but using the 
just-described conductivity. 
We choose $d_0$ to be a small fraction of the root-mean-square height $\bar{h}$ of the 
rough substrate. We varied this fraction by a factor of 1000 without a qualitative
change of the observations.

\begin{figure}[htbp]
\centering
\includegraphics[width=0.6\linewidth]{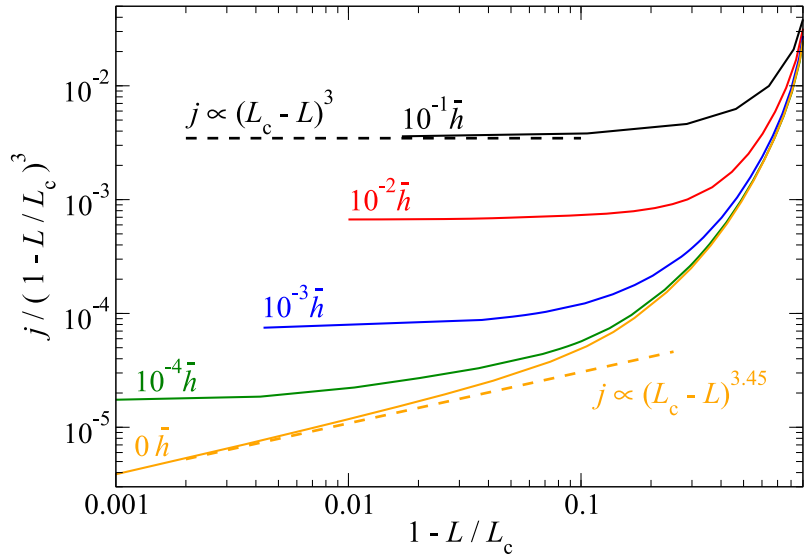}
\caption{\label{fig:slipLength}
Leakage current $j$ as a function of the reduced load $1-L/L_{\rm c}$ for a system 
with a negative slip length for the fluid flow. 
The same random realisation is studied as in Fig.~\ref{fig:defaultSystem}A--D,
where zero slip-length flow is considered. The data is smoothed to remove scatter
and shifted vertically to yield a current of one at an infinitesimally small
load.
}
\end{figure}

Figure~\ref{fig:slipLength} reveals that the flow is not affected far from
the percolation point.
However, the conductance exponent now appears to be $\beta = 3$. 
This value can be readily explained:
in the present model, the constriction (which is located around the point
where the substrate height has a saddle point) is not yet fully closed
when it appears closed for the fluid.
Thus, the point is only critical for the flow but not for the contact 
mechanics. 
This means that near the critical load $L_{\rm c}$, the true height of the
gap at the saddle point, $g_0(L)$, the true length of the constriction $l_0(L)$, 
and the true width of the constriction $w_0(L)$ are all ``simple functions'' of the
load, which each can be expanded into a Taylor series according to
$f(L) = f(L_{\rm c}) + (L-L_{\rm c}) f'(L_{\rm c})$.
The same quantities, as perceived by the fluid, e.g., the effective local height,
or the effective width of a constriction, 
have similar functional dependencies as the true height, however, different offsets. 
In fact, all offsets $f(L_{\rm c})$ for the effective quantities can be set to zero, 
since height, width, and length of the constriction --- ``as seen by the fluid'' --- 
are all zero at the critical point.
Since the resistivity 
of the constriction scales as the inverse
third power of the effective gap, linear with the length of the constriction 
(as in a serial coupling of resistors) and with the inverse width of the 
constriction (as in a parallel coupling of resistors), the fluid resistance 
of the constriction follows
\begin{equation}
R(L) \propto \frac{l_0(L)}{w_0(L) g_0^3(L)},
\label{eq:consRes}
\end{equation}
where the proportionality factor is linear in the viscosity of the
fluid and also depends on the geometry of the constriction.
Inserting our Taylor series approximations for effective height, width, and length of 
the constriction into Eq.~(\ref{eq:consRes}) then reveals that $R(L) \propto
(1-L/L_{\rm c})^{-3}$ implying $\beta = 3$. 
As discussed in a previous paper\cite{Dapp15EPL}, the case of zero slip length is 
more complicated, because $g_0(L)$, $l_0(L)$, and $w_0(L)$ all approach
zero as non-integer powerlaws of the reduced load. 

\subsection*{Critical flow through adhesive interfaces}

We now turn our attention to surfaces that attract each other via adhesive forces.
The fluid flow has stick boundary conditions again.
In a previous work\cite{Dapp15EPL}, we found, for the case of isolated 
constrictions and short-range adhesion, that constrictions closed discontinuously.
Long-range adhesion was not considered explicitly as it reduces to a simple adhesive 
offset force for the investigated single-wavelength isolated constriction (see the 
method section). 
The range of adhesion is quantified by a dimensionless number called the
Tabor coefficient $\mu_{\rm T}$\cite{Tabor77}.
Its use is best known in the context of single-asperity contacts, but the concept
extends to randomly-rough,  self-affine surfaces\cite{Persson14JCP}. 
Except for prefactors, which can be chosen at will\cite{Muser14Beil}, it is defined
as~$\mu_{\rm T} = R_{\rm c}^{1/3} (\vert \gamma_0 \vert/E^*)^{2/3}/z_0$.
Here, $z_0$ is a characteristic length scale of the interaction,
$E^*$ is the effective elastic contact modulus, and
$\gamma_0$ is the surface energy.
$R_{\rm c}$ is the radius of curvature for a Hertzian contact geometry, 
or a measure for the inverse surface curvature:
$1/R_{\rm c}^2 \equiv 
\sum_{\vec{q}} (q_x^4+q_y^4) C(\vec{q})/2$
where $C(\vec{q})$ is the (discrete) wavevector-dependent height spectrum
defined in the method section. 

\begin{figure*}[htbp]
\centering
\includegraphics[width=1.0\linewidth]{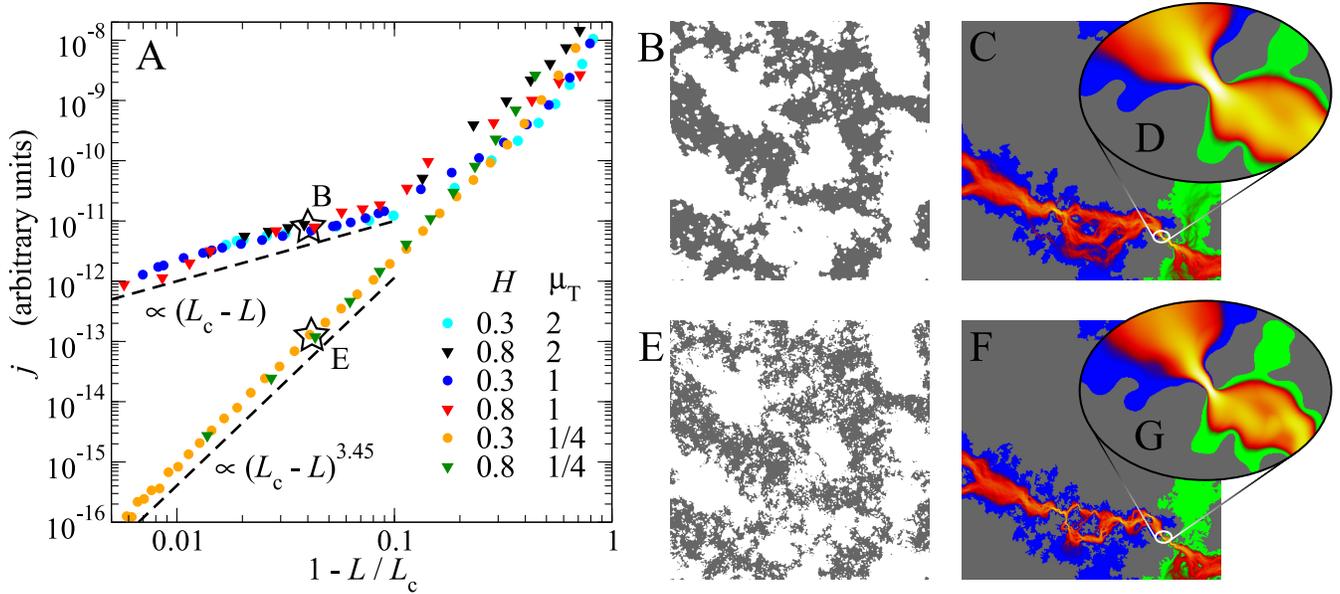}
\caption{\label{fig:adhesion}
A: Leakage current $j$ as a function of the reduced load $1-L/L_{\rm c}$ 
for adhesive contacts differing in their Hurst roughness exponent $H$ and
Tabor coefficient $\mu_{\rm T}$. 
Data is shifted vertically to superimpose in the critical regions.
B and E show the true contact near the percolation threshold for one random 
realisation with $H = 0.8$, in the case of short-range ($\mu_{\rm T} = 2$) and 
long-range ($\mu_{\rm T} = 1/4$) adhesion, respectively. 
C and F are the corresponding flow patterns at loads indicated in panel~A.
D and G are high-resolution zooms into the critical constriction. 
}
\end{figure*}

Figure~\ref{fig:adhesion} shows that the critical leakage current sensitively 
depends on the adhesive range.
Long-range adhesion yields a similar dependence of the current on the
reduced load as the non-adhesive case.
However, at a Tabor coefficient around $\mu_{\rm T}=1$, the leakage-load dependence
starts to show a different powerlaw near the sealing transition.
Specifically, for $\mu_{\rm T} \gtrsim 1$,
load regimes with an apparent conductance exponent of $\beta\approx 1$ occur.
We thus have the second example for a change of conductance exponent of macroscopic,
or at least mesoscopic, response functions due to small changes in the model.
Below, we demonstrate that the observed crossover is also present in an 
isolated constriction and thus not due to the multi-scale topology of the 
percolating channel.

An interesting aspect of Fig.~\ref{fig:adhesion} is that the coarse features
of the contact area look almost identical near the sealing transition, even 
in the two extremes of no adhesion and short-range adhesion.
However, the contact lines look much smoother and less fractal for short-range 
adhesion (E) than for no or long-range adhesion (B).
This difference in local contact features ultimately accounts for the different 
behaviour near the percolation threshold. 

\subsubsection*{Flow through isolated, adhesive constrictions}
\begin{figure}[htbp]
\centering
\includegraphics[width=0.6\linewidth]{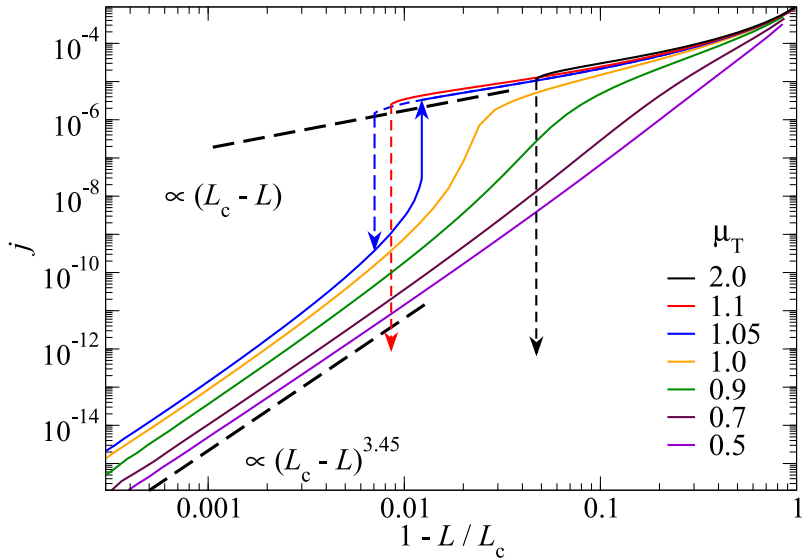}
\caption{\label{fig:isolated}
Leakage current $j$ as a function of the reduced load $1-L/L_{\rm c}$ for an 
isolated, single-wavelength constriction, for different Tabor parameters.
}
\end{figure}

We now address the question of whether the cross-over of exponents presented in 
Fig.~\ref{fig:adhesion} is due to the multi-scale roughness of the surfaces or 
originates from the properties of an isolated constriction. 
Towards this end, we revisit the contact mechanics of single-wavelength roughness, 
in particular that of a square saddle point (for details see Ref.~\citenum{Dapp15EPL}).
However, here we do not only consider short-range adhesion as in our precedent study on 
isolated constrictions, but also allow for medium- or long-range adhesion.
Figure~{\ref{fig:isolated}} shows that for $\mu_{\rm T} \lesssim 1$ the scaling of the 
current on the load changes near the 
percolation point, from the $\beta=3.45$ behaviour also seen in the non-adhesive case, 
towards a scaling with $\beta=1$.
For $\mu_{\rm T} = 1$, the latter regime is rather narrow and the leakage quickly 
becomes similar to that of non-adhesive surfaces as the sealing transition is 
approached.
In a narrow range of $\mu_{\rm T} \gtrsim 1$, the ``new'' scaling is valid
over more than one decade. For $\mu_{\rm T} \gtrsim 2$, there seems
to be a discontinuous drop of finite to zero conductance of the critical 
junction. The critical constriction snaps shut before 
scaling 
can be observed. With decreasing range of the adhesive potential, 
this point of adhesive instability is moved to smaller loads, and away from the 
critical load. 

\section*{Conclusions}

From the three leakage models analysed in this study, it has become clear that
seal systems are unlikely to belong to a (unique) universality class
of percolation theory, even if one could consider flow through a seal as a paradigm
percolation problem:
when adding small alterations to our default model, in which common 
approximations of lubrication theory are made, we observe qualitative changes
in the transition between finite and zero conductance.
In fact, it appears as though the default model represents a multi-critical point, 
because the conductance exponent changes when an arbitrarily-small negative slip 
length is introduced and/or the transition changes from continuous to discontinuous 
when short-range adhesion between the surfaces is introduced. 
In practice, many additional complications can and in general will affect the leakage
problem, most notably the formation of fluid capillaries, clogging by 
contaminating particles, as well as viscoelastic deformation and ageing of the 
sealing material.
These complications are likely to correspond to relevant perturbations affecting
the nature of the percolation transition as well. 

While it is certainly possible to predict leakage over a broad pressure range from 
statistical properties of the surfaces alone, it appears impossible to do so accurately
when the load exceeds 80\% or 90\% of the critical load, above which no open channel
percolates from one side of the interface to the other side.
At such high loads, most of the pressure pushing the fluid through the interface
drops at a single constriction. 
The (local) properties of this constriction then determine the behaviour of the
whole system. 
This is one reason why it is impossible to predict with great accuracy how strongly 
we have to tighten a water tap to make it stop dripping and also how it starts dripping
once we loosen it again.

\section*{Methods}
\subsection*{GFMD} 
The contact mechanics treatment and its description is in large parts identical 
to that presented in our study on isolated constrictions\cite{Dapp15EPL}:
We assume linear elasticity and the small-slope approximation, so that the roughness
can be mapped to a rigid substrate and the elastic compliance to a flat counter
body without loss of generality.
The effective contact modulus is used to define the unit of pressure,
i.e., $E^* = 1$.
Elasticity is treated with Green's function molecular dynamics
(GFMD)\cite{Campana06} and the continuum version of the stress-displacement
relation in Fourier space,
$\tilde{\sigma}(\vec{q}) = qE^* \tilde{u}(\vec{q})/2$,
where $\vec{q}$ is a wave vector and $q$ its magnitude.
Simulations are run in a force-controlled fashion.
After the external load has changed by a small amount, all degrees of
freedom are relaxed until convergence is attained.

Two surfaces interact with a hard-wall constraint, i.e., they are not allowed
to overlap.
In addition, we assume a finite-range surface energy 
$\gamma = -\gamma_0 \exp\{-g/z_0\}$,
where $g=g(x,y)$ is the local gap or interfacial separation.
When adhesion is switched on, we chose $\gamma_0$ such that the
contact area roughly doubles at a given load compared to the
adhesion-free case. 

The linear size of the solids is denoted by ${\cal L}$. 
Periodic boundary conditions are employed within the $xy$-plane so that
the local height in real space can be written as a Fourier sum
$h(\vec{r}) = \sum_{\vec{q}} \tilde{h}(\vec{q})
\exp(i\vec{q}\cdot\vec{r})$. 

Most technical and natural surfaces are self-affine rough, e.g.,
ground steel, asphalt, human skin, and sandblasted plexiglass show
self-similar height spectra over a broad range with Hurst roughness exponent
of $H \approx 0.8$~\cite{Persson14TL}. The exponent states that
the root-mean-square deviation of the height increases as $\Delta h \propto
\Delta r^H$, where $\Delta r$ is the in-plane distance from a given point on
the surface. An ideal random walk corresponds to $H=0.5$. A larger value of
$H$ indicates that height spectra increase in magnitude at long wavelengths
relative to short wavelengths. Sometimes $3-H$ is called the fractal
dimension of the surface.
Pertaining to the single-wavelength $\lambda$ constrictions discussed in 
Ref.~\citenum{Dapp15EPL} and in the appendix, we note that their equilibrium 
height-profile can be generated, for example, from 
$h(x,y) \equiv 2 + \cos(2\pi x/\lambda) + \cos(2\pi y/\lambda)$.

The roughness spectra $C({\bf q}) = \langle \vert \tilde{h}(\vec{q}) \vert^2 \rangle$ 
is constant for wave vectors of magnitude
$2\pi/{\cal L} \le q < 2\pi/\lambda_{\rm r}$, where
$\lambda_{\rm r}$ is called the rolloff wavelength.
For wave vectors of magnitude 
$2\pi/\lambda_{\rm r} \le q < 2\pi/\lambda_{\rm s}$,
the spectra are power laws according to $C({\bf q}) \propto q^{-2(1+H)}$. 
A typical setup can be characterised by the following dimensionless numbers:
$H = 0.8$, 
${\cal L}/\lambda_{\rm r} = 2$,
$\lambda_{\rm r}/\lambda_{\rm s} = 64$, and
$\lambda_{\rm s}/a = 64$.
When approaching the percolation threshold or when treating short-range
adhesion, we further increase the ratio
of $\lambda_{\rm s}/a$ while keeping the random realisation of the surface
profile the same. 
To determine fluctuations of the fluid conductance at small loads, we
decrease $\lambda_{\rm s}/a$ but take much larger ratios for 
${\cal L}/\lambda_{\rm r}$ and $\lambda_{\rm r}/\lambda_{\rm s}$. 
The largest GFMD calculation presented in this work consisted of
$2^{15}\times 2^{15} \approx 1\times 10^9$ discretisation points. 

\subsection*{Reynolds solver} 
For the fluid-mechanics-related calculations, we assume a reservoir of liquid
on the left side of the system ($x=0$), while the other, right side is a sink 
for said liquid, with a liquid pressure of zero.
In the transverse direction, the system is treated as periodic, in order to 
minimise finite-size effects. 
The local fluid conductivity in the Reynolds equation scales with the 
third power of the local gap as seen by the fluid.

Like the contact-mechanics aspects, all solution strategies including their
descriptions are in large parts identical to those presented in our study
on isolated constrictions\cite{Dapp15EPL}:
We use the \verb+hypre+ package\cite{FalgoutEtAl2006} to solve the sparse linear system that
the discretised Reynolds equation can be expressed as.
We employ the solvers supplied with \verb+hypre+ using the CG (conjugate
gradient), or GMRES (generalised minimal residual) methods\cite{SaadSchultz1986},
each preconditioned using the PFMG method, which is a  parallel
semicoarsening multigrid solver\cite{AshbyFalgout1996}.
Our in-house code is MPI-parallelised and uses HDF5 for I/O.

The fluid pressure and its gradients are assumed small enough to not deform 
the walls.
We verified 
that our results for the conductance exponent do not depend on this 
approximation, by including the coupling of the fluid pressure (up to 30\% of the
external mechanical pressure near the percolation threshold) to the wall for an 
isolated (adhesionless, zero-slip length) constriction. 
We merely observed an increase in 
the percolation load, a shift in the location of the critical constriction, 
and a reduction of symmetry of the gap and contact line profiles.
Coupling to GFMD is done through an iterative perturbation treatment, in which
the Reynolds output is fed back into the contact mechanics calculation. 

\section*{Acknowledgements}
We gratefully acknowledge computing time on JUROPA and JUQUEEN at the 
J\"ulich Supercomputing Centre as well as valuable discussions with Bo Persson. 
 \section*{Additional information} 
\textbf{Competing financial interests:} The authors declare no
competing financial interests.
 \section*{Author contributions}
The authors contributed to the current article as follows. M.H.M. developed the concept of the study.
W.B.D. and M.H.M. designed the contact mechanics code.
W.B.D. wrote the Reynolds solver, ran the simulations, and performed the
analysis.
W.B.D. and M.H.M. wrote the manuscript. 


\begin{thebibliography}{10}
\expandafter\ifx\csname url\endcsname\relax
  \def\url#1{\texttt{#1}}\fi
\expandafter\ifx\csname urlprefix\endcsname\relax\def\urlprefix{URL }\fi
\providecommand{\bibinfo}[2]{#2}
\providecommand{\eprint}[2][]{\url{#2}}

\bibitem{Flitney07}
\bibinfo{author}{Flitney, R.}
\newblock \emph{\bibinfo{title}{Seals and sealing handbook}}
  (\bibinfo{publisher}{Elsevier}, \bibinfo{year}{2007}).

\bibitem{Persson08JPCM}
\bibinfo{author}{Persson, B.~N.~J.} \& \bibinfo{author}{Yang, C.}
\newblock \bibinfo{title}{Theory of the leak-rate of seals}.
\newblock \emph{\bibinfo{journal}{J. Phys. Condens. Matter}}
  \textbf{\bibinfo{volume}{20}}, \bibinfo{pages}{315011}
  (\bibinfo{year}{2008}).

\bibitem{Lorenz09EPL}
\bibinfo{author}{Lorenz, B.} \& \bibinfo{author}{Persson, B.~N.~J.}
\newblock \bibinfo{title}{Leak rate of seals: Comparison of theory with
  experiment}.
\newblock \emph{\bibinfo{journal}{EPL}} \textbf{\bibinfo{volume}{86}},
  \bibinfo{pages}{44006} (\bibinfo{year}{2009}).

\bibitem{Lorenz10EPJE}
\bibinfo{author}{Lorenz, B.} \& \bibinfo{author}{Persson, B.~N.~J.}
\newblock \bibinfo{title}{Leak rate of seals: Effective-medium theory and
  comparison with experiment}.
\newblock \emph{\bibinfo{journal}{Eur. Phys. J. E}}
  \textbf{\bibinfo{volume}{31}}, \bibinfo{pages}{159} (\bibinfo{year}{2010}).

\bibitem{Persson01}
\bibinfo{author}{Persson, B. N.~J.}
\newblock \bibinfo{title}{Theory of rubber friction and contact mechanics}.
\newblock \emph{\bibinfo{journal}{J. Chem. Phys.}}
  \textbf{\bibinfo{volume}{115}}, \bibinfo{pages}{3840--3861}
  (\bibinfo{year}{2001}).

\bibitem{Persson05JPCM}
\bibinfo{author}{Persson, B.~N.~J.}, \bibinfo{author}{Albohr, O.},
  \bibinfo{author}{Tartaglino, U.}, \bibinfo{author}{Volokitin, A.~I.} \&
  \bibinfo{author}{Tosatti, E.}
\newblock \bibinfo{title}{On the nature of surface roughness with application
  to contact mechanics, sealing, rubber friction and adhesion}.
\newblock \emph{\bibinfo{journal}{J. Phys. Condens. Matter}}
  \textbf{\bibinfo{volume}{17}}, \bibinfo{pages}{R1} (\bibinfo{year}{2005}).

\bibitem{Bruggeman35}
\bibinfo{author}{Bruggeman, D.~A.~G.}
\newblock \bibinfo{title}{Calculation of various physics constants in
  heterogenous substances i - dielectricity constants and conductivity of mixed
  bodies from isotropic substances}.
\newblock \emph{\bibinfo{journal}{Ann. Phys. Lpz.}}
  \textbf{\bibinfo{volume}{24}}, \bibinfo{pages}{636} (\bibinfo{year}{1935}).

\bibitem{Reynolds86}
\bibinfo{author}{Reynolds, O.}
\newblock \bibinfo{title}{On the theory of lubrication and its application to
  \protect{Mr. Beauchamp Tower's} experiments, including an experimental
  determination of the viscosity of olive oil}.
\newblock \emph{\bibinfo{journal}{Philos. Trans. R. Soc. Lond.}}
  \textbf{\bibinfo{volume}{177}}, \bibinfo{pages}{157--234}
  (\bibinfo{year}{1886}).

\bibitem{Dapp12PRL}
\bibinfo{author}{Dapp, W.~B.}, \bibinfo{author}{L\"ucke, A.},
  \bibinfo{author}{Persson, B. N.~J.} \& \bibinfo{author}{M\"user, M.~H.}
\newblock \bibinfo{title}{Self-affine elastic contacts: percolation and
  leakage}.
\newblock \emph{\bibinfo{journal}{Phys. Rev. Lett.}}
  \textbf{\bibinfo{volume}{108}}, \bibinfo{pages}{244301}
  (\bibinfo{year}{2012}).

\bibitem{Armand64}
\bibinfo{author}{Armand, G.}, \bibinfo{author}{Lapujoulade, J.} \&
  \bibinfo{author}{Paigne, J.}
\newblock \bibinfo{title}{A theoretical and experimental relationship between
  the leakage of gases through the interface of two metals in contact and their
  superficial micro-geometry}.
\newblock \emph{\bibinfo{journal}{Vacuum}} \textbf{\bibinfo{volume}{14}},
  \bibinfo{pages}{53--57} (\bibinfo{year}{1964}).

\bibitem{Kirkpatrick73}
\bibinfo{author}{Kirkpatrick, S.}
\newblock \bibinfo{title}{Percolation and conduction}.
\newblock \emph{\bibinfo{journal}{Rev. Mod. Phys.}}
  \textbf{\bibinfo{volume}{45}}, \bibinfo{pages}{574--588}
  (\bibinfo{year}{1973}).

\bibitem{Bottiglione09}
\bibinfo{author}{Bottiglione, F.}, \bibinfo{author}{Carbone, G.},
  \bibinfo{author}{Mangialardi, L.} \& \bibinfo{author}{Mantriota, G.}
\newblock \bibinfo{title}{Leakage mechanism in flat seals}.
\newblock \emph{\bibinfo{journal}{J. Appl. Phys.}}
  \textbf{\bibinfo{volume}{106}}, \bibinfo{pages}{104902}
  (\bibinfo{year}{2009}).

\bibitem{Bottiglione09TI}
\bibinfo{author}{Bottiglione, F.}, \bibinfo{author}{Carbone, G.} \&
  \bibinfo{author}{Mantriota, G.}
\newblock \bibinfo{title}{Fluid leakage in seals: An approach based on
  percolation theory}.
\newblock \emph{\bibinfo{journal}{Tribol. Int.}} \textbf{\bibinfo{volume}{42}},
  \bibinfo{pages}{731--737} (\bibinfo{year}{2009}).

\bibitem{Dapp15EPL}
\bibinfo{author}{Dapp, W.~B.} \& \bibinfo{author}{M\"user, M.~H.}
\newblock \bibinfo{title}{Contact mechanics of and reynolds flow through saddle
  points}.
\newblock \emph{\bibinfo{journal}{EPL}} \textbf{\bibinfo{volume}{109}},
  \bibinfo{pages}{44001} (\bibinfo{year}{2015}).

\bibitem{Vallet09a}
\bibinfo{author}{Vallet, C.}, \bibinfo{author}{Lasseux, D.},
  \bibinfo{author}{Sainsot, P.} \& \bibinfo{author}{Zahouani, H.}
\newblock \bibinfo{title}{Real versus synthesized fractal surfaces: Contact
  mechanics and transport properties}.
\newblock \emph{\bibinfo{journal}{Tribol. Int.}} \textbf{\bibinfo{volume}{42}},
  \bibinfo{pages}{250--259} (\bibinfo{year}{2009}).

\bibitem{Vallet09b}
\bibinfo{author}{Vallet, C.}, \bibinfo{author}{Lasseux, D.},
  \bibinfo{author}{Zahouani, H.} \& \bibinfo{author}{Sainsot, P.}
\newblock \bibinfo{title}{Samping effect on contact and transport properties
  between fractal surfaces}.
\newblock \emph{\bibinfo{journal}{Tribol. Int.}} \textbf{\bibinfo{volume}{42}},
  \bibinfo{pages}{1132--1145} (\bibinfo{year}{2009}).

\bibitem{Stauffer91}
\bibinfo{author}{Stauffer, D.} \& \bibinfo{author}{Aharony, A.}
\newblock \emph{\bibinfo{title}{An Introduction to Percolation Theory}}
  (\bibinfo{publisher}{CRC Press}, \bibinfo{year}{1991}).

\bibitem{Webman77}
\bibinfo{author}{Webman, I.}, \bibinfo{author}{Jortner, J.} \&
  \bibinfo{author}{Cohen, M.~H.}
\newblock \bibinfo{title}{Critical exponents for percolation conductivity in
  resistor networks}.
\newblock \emph{\bibinfo{journal}{Phys. Rev. B}} \textbf{\bibinfo{volume}{16}},
  \bibinfo{pages}{2593--2596} (\bibinfo{year}{1977}).

\bibitem{Schrenk13}
\bibinfo{author}{Schrenk, K.~J.} \emph{et~al.}
\newblock \bibinfo{title}{Percolation with long-range correlated disorder}.
\newblock \emph{\bibinfo{journal}{Phys. Rev. E}} \textbf{\bibinfo{volume}{88}},
  \bibinfo{pages}{052102} (\bibinfo{year}{2013}).

\bibitem{Grassberger13}
\bibinfo{author}{Grassberger, P.}
\newblock \bibinfo{title}{Two-dimensional \protect{SIR} epidemics with
  long-range infection}.
\newblock \emph{\bibinfo{journal}{J. Stat. Phys.}}
  \textbf{\bibinfo{volume}{153}}, \bibinfo{pages}{289--311}
  (\bibinfo{year}{2013}).

\bibitem{Almqvist11}
\bibinfo{author}{Almqvist, A.}, \bibinfo{author}{Campa{\~n}\'a, C.},
  \bibinfo{author}{Prodanov, N.} \& \bibinfo{author}{Persson, B. N.~J.}
\newblock \bibinfo{title}{Interfacial separation between elastic solids with
  randomly rough surfaces: Comparison between theory and numerical techniques}.
\newblock \emph{\bibinfo{journal}{J. Mech. Phys. Solids}}
  \textbf{\bibinfo{volume}{59}}, \bibinfo{pages}{2355} (\bibinfo{year}{2011}).

\bibitem{SchnyderEtAl2015}
\bibinfo{author}{Schnyder, S.~K.}, \bibinfo{author}{Spanner, M.},
  \bibinfo{author}{Hofling, F.}, \bibinfo{author}{Franosch, T.} \&
  \bibinfo{author}{Horbach, J.}
\newblock \bibinfo{title}{Rounding of the localization transition in model
  porous media}.
\newblock \emph{\bibinfo{journal}{Soft Matter}} \textbf{\bibinfo{volume}{11}},
  \bibinfo{pages}{701--711} (\bibinfo{year}{2015}).

\bibitem{Tabor77}
\bibinfo{author}{Tabor, D.}
\newblock \bibinfo{title}{Surface forces and surface interactions}.
\newblock \emph{\bibinfo{journal}{J. Colloid Interface Sci.}}
  \textbf{\bibinfo{volume}{58}}, \bibinfo{pages}{2--13} (\bibinfo{year}{1977}).

\bibitem{Persson14JCP}
\bibinfo{author}{Persson, B. N.~J.} \& \bibinfo{author}{Scaraggi, M.}
\newblock \bibinfo{title}{Theory of adhesion: Role of surface roughness}.
\newblock \emph{\bibinfo{journal}{J. Chem. Phys.}}
  \textbf{\bibinfo{volume}{141}}, \bibinfo{pages}{124701}
  (\bibinfo{year}{2014}).

\bibitem{Muser14Beil}
\bibinfo{author}{M\"user, M.~H.}
\newblock \bibinfo{title}{Single-asperity contact mechanics with positive and
  negative work of adhesion}.
\newblock \emph{\bibinfo{journal}{Beilstein J. Nanotechnol.}}
  \textbf{\bibinfo{volume}{5}}, \bibinfo{pages}{419--437}
  (\bibinfo{year}{2014}).

\bibitem{Campana06}
\bibinfo{author}{Campa{\~n}\'a, C.} \& \bibinfo{author}{M\"user, M.~H.}
\newblock \bibinfo{title}{Practical green's function approach to the simulation
  of elastic semi-infinite solids}.
\newblock \emph{\bibinfo{journal}{Phys. Rev. B}} \textbf{\bibinfo{volume}{74}},
  \bibinfo{pages}{075420} (\bibinfo{year}{2006}).

\bibitem{Persson14TL}
\bibinfo{author}{Persson, B. N.~J.}
\newblock \bibinfo{title}{On the fractal dimension of rough surfaces}.
\newblock \emph{\bibinfo{journal}{Tribol. Lett.}}
  \textbf{\bibinfo{volume}{54}}, \bibinfo{pages}{99--106}
  (\bibinfo{year}{2014}).

\bibitem{FalgoutEtAl2006}
\bibinfo{author}{Falgout, R.~D.}, \bibinfo{author}{Jones, J.~E.} \&
  \bibinfo{author}{Meier~Yang, U.}
\newblock \bibinfo{title}{The design and implementation of hypre, a library of
  parallel high performance preconditioners}.
\newblock In \bibinfo{editor}{Bruaset, A.} \& \bibinfo{editor}{Tveito, A.}
  (eds.) \emph{\bibinfo{booktitle}{Numerical Solution of Partial Differential
  Equations on Parallel Computers}}, \bibinfo{number}{51},
  \bibinfo{pages}{267--294} (\bibinfo{publisher}{Springer-Verlag},
  \bibinfo{year}{2006}).

\bibitem{SaadSchultz1986}
\bibinfo{author}{Saad, Y.} \& \bibinfo{author}{Schultz, M.~H.}
\newblock \bibinfo{title}{\protect{GMRES}: A generalized minimal residual algorithm for
  solving nonsymmetric linear systems}.
\newblock \emph{\bibinfo{journal}{SIAM J. Sci. Stat. Comput.}}
  \textbf{\bibinfo{volume}{7}}, \bibinfo{pages}{856--869}
  (\bibinfo{year}{1986}).

\bibitem{AshbyFalgout1996}
\bibinfo{author}{Ashby, S.~F.} \& \bibinfo{author}{Falgout, R.~D.}
\newblock \bibinfo{title}{A parallel multigrid preconditioned conjugate
  gradient algorithm for groundwater flow simulations}.
\newblock \emph{\bibinfo{journal}{Nucl. Sci. Eng.}}
  \textbf{\bibinfo{volume}{124}}, \bibinfo{pages}{145--159}
  (\bibinfo{year}{1996}).

\end{thebibliography}

\end{document}